\documentclass{camera}
\usepackage{graphicx}  % uncomment this if your want to insert figures
\usepackage{cite} % more options in command \cite
\usepackage{verbatim}

\begin{document}

%%%%%%%%%%%%%%%%%%%%%%%%%%%%%%%%%%%%%%%%%%%%%%%%%%%%%%%%
% The title, only the first letter capitalized; if you want to split it in
% two or more lines, put a \\ macro at each line break
% example: 
%   \title{Title: first line\\ second line}
%
\title{Search for the $\eta$-mesic Helium bound state with the WASA-at-COSY facility}

\author{Magdalena Skurzok$^{a}$, Wojciech Krzemie{\'n}$^{b}$, Oleksandr Rundel$^{a}$ \and Pawel Moskal$^{a}$ \\
for the WASA-at-COSY Collaboration}

%%%%%%%%%%%%%%%%%%%%%%%%%%%%%%%%%%%%%%%%%%%%%%%%%%%%%%%%
%
\organization{$^{a}$ M. Smoluchowski Institute of Physics, Jagiellonian University, Cracow, Poland\\
$^{b}$ National Centre of Nuclear Research, {\'S}wierk, Poland}

\maketitle

\begin{abstract}

We performed a search for $^{4}\hspace{-0.03cm}\mbox{He}$-$\eta$ bound state with high statistics and high acceptance with the WASA-at-COSY facility using a ramped beam technique. The signature of $\eta$-mesic nuclei is searched for in $dd\rightarrow$ $^{3}\hspace{-0.03cm}\mbox{He} n \pi{}^{0}$ and $dd\rightarrow$ $^{3}\hspace{-0.03cm}\mbox{He} p \pi{}^{-}$ reactions by the measurement of the excitation functions in the vicinity of the $\eta$ production threshold.~This paper presents the experimental method and the preliminary results of the data analysis for $dd\rightarrow$ $^{3}\hspace{-0.03cm}\mbox{He} n \pi{}^{0}$ process.

\end{abstract}

%%%%%%%%%%%%%%%%%%%%%%%%%%%%%%%%%%%%%%%%%%%%%%%%%%%%%%%%
% Write the text starting from here and using the usual
% LaTeX commands.
%
\section{Introduction}

In 1986 Haider and Liu~\cite{HaiderLiu1}, based on the coupled-channel analysis of the $\pi N \rightarrow \pi N$ , $\pi N \rightarrow \pi \pi N$ and $\pi N \rightarrow \eta N$ reactions~\cite{BhaleraoLiu}, postulated the existence of $\eta$-mesic nuclei in which the $\eta$ meson is bound within a nucleus via the strong interaction. Since then, the search for $\eta$- and $\eta'$-mesic bound states was performed in many laboratories as e.g.: COSY~\cite{MoskalSmyrski, MSkurzok, Adlarson_2013,Krzemien_PhD,Budzanowski, Mersmann, Smyrski1}, ELSA~\cite{Nanova2}, GSI~\cite{Tanaka}, JINR~\cite{Afanasiev}, JPARC~\cite{Fujioka}, LPI~\cite{Baskov}, and MAMI~\cite{Krusche_2013, Pheron}, however, till now none of the experiments confirmed its existence. The status of the search was recently described in the following reviews~\cite{Machner_2015,Kelkar,Krusche_Wilkin}.

\indent Recent theoretical studies e.g.~\cite{BassTom1, WycechKrzemien, Hirenzaki_2010, Hirenzaki1, Friedman_2013, Wilkin2, Nagahiro_2013, Kelkar} support the search for $\eta$ and $\eta'$-mesic nuclei. Phenomenological and theoretical investigations of $\eta$ production in hadronic- and photo- induced reactions result in a wide range of possible scattering lengths from $a_{\eta N}=(0.18+0.16i)$~fm to $a_{\eta N}=(1.03+0.49i)$~fm~\cite{Kelkar}, which do not exclude the formation of a bound states for the helium~\cite{Wilkin1,WycechGreen} and even the deuteron~\cite{Green}. The observation of an steep rise in the total cross-section of the \mbox{$dp\rightarrow$ $^{3}\hspace{-0.03cm}\mbox{He}\eta$} and $dd\rightarrow$ $^{4}\hspace{-0.03cm}\mbox{He}\eta$ reactions close to the kinematic threshold~\cite{Berger,Mayer,Smyrski1,Mersmann,Frascaria,Willis,Machner_ActaPhys,Wronska}, gives a strong evidence for the existence of a bound state. Similar observations were carried out in case of the cross section for the photoproduction process $\gamma^{3}\hspace{-0.03cm}\mbox{He}\rightarrow$ $^{3}\hspace{-0.03cm}\mbox{He}\eta$~\cite{Pfeiffer,Pheron}. It shows that the rise of the cross section above threshold is independent of the initial channel giving a strong argument for the existence of the pole in the scattering matrix which could be associated with a bound state. Therefore, we consider the $^{4}\hspace{-0.03cm}\mbox{He}$-$\eta$ and $^{3}\hspace{-0.03cm}\mbox{He}$-$\eta$ as a good candidates for the $\eta$-mesic bound states.

\indent The discovery of postulated $\eta$-mesic nuclei would be interesting on its own as well as would be very important for (i) better understanding of the $\eta$ and $\eta'$ meson properties~\cite{InoueOset} and their interaction with nucleons inside nuclear matter, (ii) providing information about the $N^{*}$(1535) resonance~\cite{Hirenzaki_2010,Jido} and (vi) about flavour singlet component of the quark-gluon wave function of the $\eta$ and $\eta'$ mesons~\cite{BassTom,BassTomek2}.

\section{Search for $^{4}\hspace{-0.03cm}\mbox{He}$-$\eta$ bound state with WASA}

We performed the search for the $^{4}\hspace{-0.03cm}\mbox{He}$-$\eta$ bound state with unique accuracy with the WASA facility, installed at the COSY synchrotron in Forschungszentrum J\"ulich (Germany). The advantage of this detection system is the possibility of a simultaneous measurement of all ejectiles with high acceptance while continuously changing the beam momentum around the $\eta$ production threshold. The signature of the \mbox{$\eta$-mesic} nuclei is searched for by studying the excitation function for the chosen decay channels of the \mbox{$^{4}\hspace{-0.03cm}\mbox{He}$-$\eta$} system, formed in deuteron-deuteron collision~\cite{Moskal1,Krzemien_PhD}.~Till now two experiments were carried out focusing on the bound state decay into $^{3}\hspace{-0.03cm}\mbox{He}$ and a nucleon-pion pair~\cite{MSkurzok,Krzemien_PhD,WKrzemien_2014,Adlarson_2013,MSkurzok_PhD}. 

\indent The first experiment was carried out in June 2008, by measuring the excitation function of the $dd\rightarrow$ $^{3}\hspace{-0.03cm}\mbox{He} p \pi{}^{-}$ reaction near the $\eta$ production threshold covering the excess energy range from -51.4~MeV up to 22 MeV. In the excitation function no structure which could be interpreted as a resonance originating from decay of the $\eta$-mesic $^{4}\hspace{-0.03cm}\mbox{He}$ was observed. The upper limit of the total cross-section for the bound state formation and decay in the \mbox{$dd \rightarrow$ ($^{4}\hspace{-0.03cm}\mbox{He}$-$\eta)_{bound} \rightarrow$ $^{3}\hspace{-0.03cm}\mbox{He} p \pi{}^{-}$} process was determined and varies from 20 nb to 27 nb at the confidence level 90\%~\cite{WKrzemien_2014,Adlarson_2013}. 

\indent In the second experiment, in November 2010, about 10 times higher statistics were collected with respect to the previous measurement.~The search for the $^{4}\hspace{-0.03cm}\mbox{He}$-$\eta$ bound state was performed for two reactions $dd\rightarrow$ $^{3}\hspace{-0.03cm}\mbox{He} n \pi{}^{0}$ and $dd\rightarrow$ $^{3}\hspace{-0.03cm}\mbox{He} p \pi{}^{-}$ via the measurement of the excitation function for each of them in the vicinity of the $\eta$ production threshold. The deuteron beam momentum was varying continuously within each acceleration cycle from 2.127~GeV/c to 2.422~GeV/c, crossing the kinematic threshold for the $\eta$ production in the $dd \rightarrow$ $^{4}\hspace{-0.03cm}\mbox{He}\eta$ reaction at 2.336~GeV/c, which corresponds to a range of excess energy \mbox{Q$\in$(-70,30)~MeV}. 

\indent Here we show result for the channel $dd\rightarrow$ $^{3}\hspace{-0.03cm}\mbox{He} n \pi{}^{0}$~\cite{MSkurzok_PhD}.
% starting with the identification of all measured particles based on Monte Carlo simulations for the $\eta$-mesic nuclei production and decay. The process is schematically presented in Fig.~\ref{free_reaction}.
%\begin{figure}[h!]
%\centering
%\includegraphics[width=11.5cm,height=5.5cm]{Plots/dd_BS_3Henpi0_kinematics.png}
%\caption{Scheme of the $^{4}\hspace{-0.03cm}\mbox{He}$-$\eta$ bound state production and decay in \mbox{$dd\rightarrow$ $^{3}\hspace{-0.03cm}\mbox{He} n \pi{}^{0}$} reaction.\label{free_reaction}}
%\end{figure}

%According to the scheme, the deuteron-deuteron collision leads to the formation of $^{4}\hspace{-0.03cm}\mbox{He}$ nucleus bound with the $\eta$ meson via the strong interaction. Then, the $\eta$ meson can be absorbed by one of the neutrons inside helium and may propagate in the nucleus via the consecutive excitation of neutrons to the $N^{*}(1525)$ resonance until its decay into the neutron-pion pair. The $^{3}\hspace{-0.03cm}\mbox{He}$ plays the role of a spectator which according to the momentum conservation in the $^{4}\hspace{-0.03cm}\mbox{He}$ system moves with the Fermi momentum in the opposite direction to the $N^{*}$ resonance. The description of the bound state production and decay kinematics is presented in details in Ref.~\cite{Skurzok_Master}.

\indent The $^{3}\hspace{-0.03cm}\mbox{He}$ was identified in the Forward Detector based on the \mbox{$\Delta$E-E method}. The neutral pion $\pi^{0}$ was reconstructed in the Central Detector from the invariant mass of two gamma quanta originating from its decay while the neutron was identified via the missing mass technique. 
%The appropriate spectra with applied cuts are presented in Fig.~\ref{pion_ident}.
%\vspace{-0.3cm}
%
%\begin{figure}[h!]
%\centering
%\includegraphics[width=6.0cm,height=4.0cm]{Plots/IM_pion_lev2_cut2_1_VetoFRH2_new.png}
%\includegraphics[width=6.0cm,height=4.0cm]{Plots/MM_neutron_lev2_cut2_VetoFRH2.png}
%\vspace{-0.3cm}
%\caption{(left panel) $\pi^{0}$ and neutron identification via cut in the invariant mass spectrum and (right panel) the cut in the missing mass spectrum, respectively. Applied cuts are marked with green lines.~\label{pion_ident}}  
%\end{figure}

Events corresponding to the bound states production were selected by applying cuts in the $^{3}\hspace{-0.03cm}\mbox{He}$ center of mass (CM) momentum, nucleon CM kinetic energy, pion CM kinetic energy and the opening angle between neutron-pion pair in the CM based on Monte Carlo simulations.
%(See e.g. Fig.~\ref{fig2}). 

%%%%%%%%%%%%%%%%%%%%%%%%%%%%%%%%%%%%%%%%%%%%%%%%%%%%%%%%%%%%%%%%%%%%%%%%%%%%%%%%%%%%%%

%\begin{figure}[h!]
%\centering
%\includegraphics[width=6.0cm,height=4.0cm]{p_3He_cm_comparison_lev2_cut3_DATA_Magda_WMCSignal_region4_VetoFRH2.png} 
%\includegraphics[width=6.0cm,height=4.0cm]{Ekin_nucleon_lev7_cut1_TEST2.png}\\
%\includegraphics[width=6.0cm,height=4.0cm]{Plots/Ekin_pion_lev3_cut2_TEST2.png} 
%\includegraphics[width=6.0cm,height=4.0cm]{Plots/OA_lev7_cut3_TEST2.png}
%\vspace{-0.2cm}
%\caption{Spectrum of $p^{cm}_{^{3}\hspace{-0.05cm}He}$ (left panel), $E^{cm}_{kin_{n}}$ (right panel), 
%$E^{cm}_{kin_{\pi^{0}}}$ (left lower panel) and $\theta^{cm}_{n,\pi^{0}}$ (right lower panel). 
%Data for $dd\rightarrow$ $^{3}\hspace{-0.03cm}\mbox{He} n \pi{}^{0}$ reaction are shown in red. Monte Carlo simulations of the signal are marked by black line while the applied cuts are by the green lines. The figure is adopted from~\cite{MSkurzok_PhD}.~\label{fig2}}
%\end{figure}

\begin{figure}[h!]
\centering
\includegraphics[width=6.0cm,height=4.0cm]{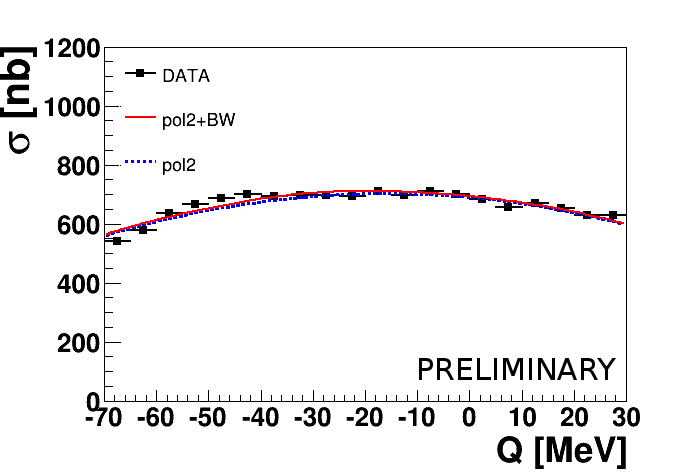}
\includegraphics[width=6.0cm,height=4.0cm]{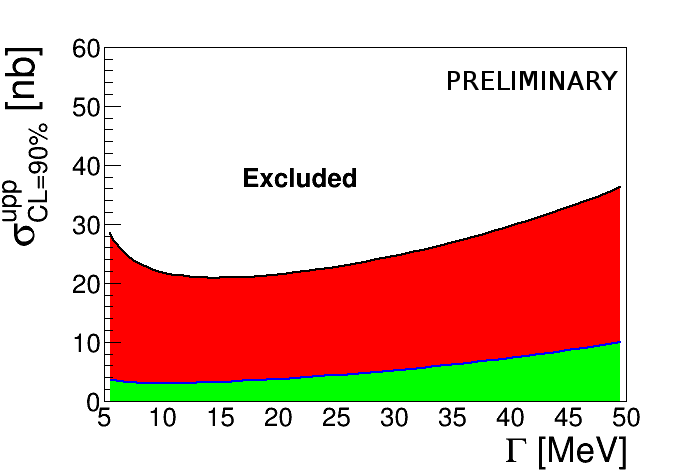}
\vspace{-0.2cm}
\caption{
(Left) Excitation function for the $dd\rightarrow$ $^{3}\hspace{-0.03cm}\mbox{He} n \pi{}^{0}$ reaction. The red solid line represents a fit with second order polynomial combined with a Breit-Wigner function with fixed binding energy and width equal to 30 and 40~MeV, respectively. The blue dotted line shows the second order polynomial corresponding to the background. The figure is adopted from~\cite{MSkurzok_PhD}.
(Right) Upper limit of the total cross-section for $dd\rightarrow(^{4}\hspace{-0.03cm}\mbox{He}$-$\eta)_{bound}\rightarrow$ $^{3}\hspace{-0.03cm}\mbox{He} n \pi{}^{0}$ reaction as a function of the width of the bound state. The binding energy was set to 30~MeV. The green area denotes the systematic uncertainties. The figure is adopted from~\cite{MSkurzok_PhD} 
\label{aa}
}
\end{figure}

The excitation functions for the reaction were determined for a region rich in signal corresponding to momenta of the $^{3}\hspace{-0.03cm}\mbox{He}$ in the CM system in range \mbox{$p^{cm}_{^{3}\hspace{-0.05cm}He}\in(0.1,0.2)$~GeV/c}.
% (region A in the left panel in Fig.~\ref{fig2}). 
The excitation function was obtained by normalizing the events selected in individual excess energy intervals by the corresponding integrated luminosities (the detailed description of the luminosity determination one can find in Ref.~\cite{MSkurzok_PhD,MSkurzok_2015}) and corrected for acceptance and efficiency is presented in the left panel of Fig.~\ref{aa}.

No narrow resonance-like strucutre, which could be interpreted as the indication of the $\eta$-mesic nuclei, is observed~\cite{MSkurzok_PhD}. Therefore, the upper limit of the total cross section for the $dd\rightarrow(^{4}\hspace{-0.03cm}\mbox{He}$-$\eta)_{bound}\rightarrow$ $^{3}\hspace{-0.03cm}\mbox{He} n \pi{}^{0}$ process was determined on the 90\% confidence level. For this purpose the excitation function was fitted with a quadratic function describing the background combined with the Breit-Wigner function which can describe the signal from the bound state. In performing the fit, the binding energy and the bound state width are fixed parameters while the polynomial coefficients and the amplitude of the Breit-Wigner distribution are treated as free parameters. The fitting procedure was performed for various values of binding energy and width. An example of the fit for $\Gamma=40$~MeV and $B_{s}=30$~MeV is presented in Fig.~\ref{aa}. The upper limit of the total cross section for considered process varies from 21 to 36~nb for the bound state width ranging from 5 to 50~MeV, while the systematic error varies from 12\% to 27\% (See Fig.~\ref{aa}).

\section{Conclusion}

The performed analysis allows for the determination of the excitation function for $dd\rightarrow(^{4}\hspace{-0.03cm}\mbox{He}$-$\eta)_{bound}\rightarrow$ $^{3}\hspace{-0.03cm}\mbox{He} n \pi{}^{0}$ process and the estimation of the upper limit of the cross section for the $\eta$-mesic $^{4}\hspace{-0.03cm}\mbox{He}$ formation and decay. The obtained excitation function does not show any narrow structure which could be a signature of the bound state with width less than 50~MeV. The upper limit is by a factor of five larger than theoretically predicted value~\cite{WycechKrzemien} therefore, the current measurement does not exclude the existence of bound state in this process.

\section{Acknowledgements}

\noindent We acknowledge support by the Foundation for Polish Science - MPD program, co-financed by the European
Union within the European Regional Development Fund, by the Polish National Science Center through grants No.~2013/11/N/ST2/04152, 2011/01/B/ST2/00431, 2011/03/B/ST2/01847 and by the FFE grants of the Forschungszentrum J\"ulich.

\addcontentsline{toc}{chapter}{Bibliography}

\thispagestyle{plain}

\end{document}